\documentclass[aps,prl,showpacs,preprint]{revtex4-1}

\usepackage{amsmath} % math extension
\usepackage{amssymb} % symbols
\usepackage{graphicx}% include figure files
\usepackage{color}

\begin{document}

\title{Time-Dependent Density Functional Theory
for Driven Lattice Gas Systems with Interactions}

\author{Marcel Dierl$^{1,2}$}
\author{Philipp Maass$^2$}
\author{Mario Einax$^{1,2}$}
\affiliation{
$^{1}$Institut f\"ur Physik, Technische Universit\"at Ilmenau,
    98684 Ilmenau, Germany\\
$^{2}$
Fachbereich Physik, Universit\"at Osnabr\"uck,
      Barbarastra\ss e 7, 49076 Osnabr\"uck, Germany}

\date{\today}

\begin{abstract}

  We present a new method to describe the kinetics of driven lattice
  gases with particle-particle interactions beyond hard-core
  exclusions. The method is based on the time-dependent density
  functional theory for lattice systems and allows one to set up
  closed evolution equations for mean site occupation numbers in a
  systematic manner. Application of the method to a totally asymmetric
  site exclusion process with nearest-neighbor interactions yields
  predictions for the current-density relation in the bulk, the phase
  diagram of non-equilibrium steady states and the time evolution of
  density profiles that are in good agreement with results from
  kinetic Monte Carlo simulations.

\end{abstract}

% 05.50.+q Lattice theory and statistics (Ising, Potts, etc.)
% 05.60.Cd Classical transport
% 05.70.Ln Nonequilibrium and irreversible thermodynamics
% 31.15.E- Density functional theory

\maketitle

Driven lattice gases are an active topic in non-equilibrium
statistical mechanics due to both their manifold applications and
their importance in fundamental studies of non-equilibrium systems
(for reviews, see
\cite{Schmittmann:1995,Derrida:1998,Schuetz:2001,Golinelli:2006,Blythe:2007}). Prominent
examples for applications to non-equilibrium processes in nature are
biopolymerization \cite{MacDonald:1968}, unidirectional motions of
motor proteins along filaments or microtubuli
\cite{Lipowsky:2001,Nishinari:2005}, flow of molecules through vessels
or ion conduction through membrane channels
\cite{Hille:2001,Einax:2010}, incoherent transport of electrons in
molecular wires \cite{Joachim:2005}, surface growth \cite{Spohn:1991}
and traffic \cite{Helbing:2001}. With respect to fundamental aspects,
questions pertaining to the theoretical description of
boundary-induced phase transitions and of non-equilibrium steady
states (NESS) can be studied within a conceptually simple framework
\cite{Derrida:1998,Schuetz:2001}.

A standard model of a driven lattice gas system is the asymmetric
simple exclusion process (ASEP), which refers to the directional
stochastic hopping transport of particles on a one-dimensional lattice
with hard-core exclusions that prevent multiple occupation of a
lattice site. In an open system with particle
injections and ejections at the left and right boundaries, this model
exhibits phase transitions between non-equilibrium steady states of
different mean site occupation in the bulk
\cite{Krug:1991,Derrida:1992}.  Analytical methods like the matrix
approach or Bethe ansatz have been developed and successfully applied
to calculate exactly the distribution of microstates in the NESS
\cite{Derrida:1998,Schuetz:2001,Golinelli:2006,Blythe:2007}. Various
extensions of the ASEP have been considered in the past, as, for
example, particle rods covering several lattice sites
\cite{Lakatos:2003}, or stochastic injection/ejection (Langmuir
kinetics) of particles in the bulk \cite{Frey:2003}. So far, however,
comparatively few studies exist \cite{Katz:1984,Krug:1991,Hager:2001}
which are concerned with the description of driven lattice gas
systems, where the particles do not only interact via athermal
hard-core exclusion.

In this work we will focus on developing a method for treating the
kinetics of such systems with interactions beyond hard-core
exclusions. In general, starting with the master equation for the
Markovian time evolution of the microstates, exact evolution equations
can be derived for the mean site occupation numbers (henceforth called
densities). These have the form of a continuity equation with currents
depending on equal-time correlations of occupation numbers (henceforth
called ``correlators''). The challenge is to develop proper methods to
calculate these correlators so that closed evolution equations for the
densities result that well account for the kinetic behavior.  In the
treatment of interaction effects in the driven system considered by
Sch\"utz {\it et al.} \cite{Hager:2001} this problem was addressed by
utilizing the special jump rates introduced in the Katz-Lebowitz-Spohn
model \cite{Katz:1984}, which yield a site occupation statistics in a
driven bulk system that can be exactly mapped onto an equilibrium
system with nearest-neighbor interactions (Ising system). As a
consequence, the current-density relation in the bulk could be
determined exactly and, by employing the maximum and minimum current
principle \cite{Krug:1991,Hager:2001}, the phase diagram of NESS in an
open system calculated in good agreement with Monte Carlo simulations.

In this Letter we show how by using the time-dependent density
functional theory (TDFT) for lattice systems a systematic method is
provided that allows one to treat the kinetics of driven lattice gases
with interactions. To demonstrate the procedure, we consider a totally
asymmetric site exclusion process (TASEP) with nearest-neighbor
interactions. Predictions of the theory for the current-density
relation in the bulk, the phase diagram of NESS, the density profiles
in the NESS, and the time evolution of density profiles are compared
to results of kinetic Monte Carlo (KMC) simulation and yield a surprisingly
good agreement for this far-from-equilibrium system.

The Hamiltonian for a one-dimensional lattice gas with
nearest-neighbor interactions $V$ is
\begin{equation}
  \mathcal{H}\left(\boldsymbol{n}\right)
= V\sum_{i} n_i n_{i+1}\,.
  \label{eq:Hamiltonian}
\end{equation}
The set of occupation numbers $\boldsymbol{n} = \left\{n_i\right\}$
specifies the microstate of the system, where $n_i = 0$ or $1$ if site
$i$ is vacant or occupied by a particle, respectively ($n_i^2=n_i$).
For unidirectional nearest-neighbor hopping considered in the TASEP,
the stochastic dynamics of the system is specified by the rates
$\Gamma_i(\boldsymbol{n})$ for a particle on site $i$ to jump to
a vacant neighboring site $(i+1)$ in the configuration
$\boldsymbol{n}$. Starting from the master equation for the time
evolution of the probability density $P(\boldsymbol{n},t)$ of
microstates, one derives the discrete version of the continuity
equation for the densities $ \rho_i(t)=\langle n_i\rangle_t$ (see,
e.~g.\ \cite{Gouyet:2003} for a general derivation),
\begin{equation}
  \frac{d\rho_i(t)}{dt}= j_{i-1}\left(t\right) - j_i\left(t\right)\,,
  \label{eq:rate_eq}
\end{equation}
where the mean currents from $i$ to $i+1$ are given by
\begin{equation}
  j_i(t) = \left\langle n_i\left(1 -
  n_{i+1}\right)\Gamma_i(\boldsymbol{n})\right\rangle_t\,,
  \label{eq:current}
\end{equation}
and $\left\langle\cdots\right\rangle_t$ denotes an average with
respect to $P(\boldsymbol{n},t)$. To complete the specification of the
dynamics we define the jump rates $\Gamma_i(\boldsymbol{n})$ as \cite{comm:rates}
\begin{equation}
\Gamma_i(\boldsymbol{n})=
\nu\exp\left[-\left(\mathcal{H}(\boldsymbol{n}^{(i,i+1)})-
\mathcal{H}(\boldsymbol{n})\right)/2k_\text{B}T\right]\,.
  \label{eq:rate}
\end{equation}
Here $\nu$ is an attempt frequency, and $\boldsymbol{n}^{(i,i+1)}$
refers to the target configuration of the jump, where, with respect to
the initial configuration $\boldsymbol{n}$, $n_i$ and $n_{i+1}$ are
interchanged, while the other $n_k$ are the same. We use $\nu^{-1}$ as
time unit and $k_\text{B}T$ as energy unit in the following
($\nu^{-1}=1$ and $k_\text{B}T=1$). With Eq.~(\ref{eq:rate}) the
currents can be written explicitely as
\begin{align}
j_i&=\langle \tilde n_{i-1}n_i\tilde n_{i+1}\tilde n_{i+2}\rangle_t
+e^{V/2}\langle n_{i-1}n_i\tilde n_{i+1}\tilde n_{i+2}\rangle_t+e^{-V/2}\langle \tilde n_{i-1}n_i\tilde n_{i+1}n_{i+2}\rangle_t
+\langle n_{i-1}n_i\tilde n_{i+1}n_{i+2}\rangle_t\,,
\label{eq:current2}
\end{align}
where we introduced the hole occupation numbers $\tilde n_i=1-n_i$.

%------ FIGURE 1 ------------------------------------------------
\begin{figure}[H!]
\centering
 \includegraphics[width=0.6\textwidth]{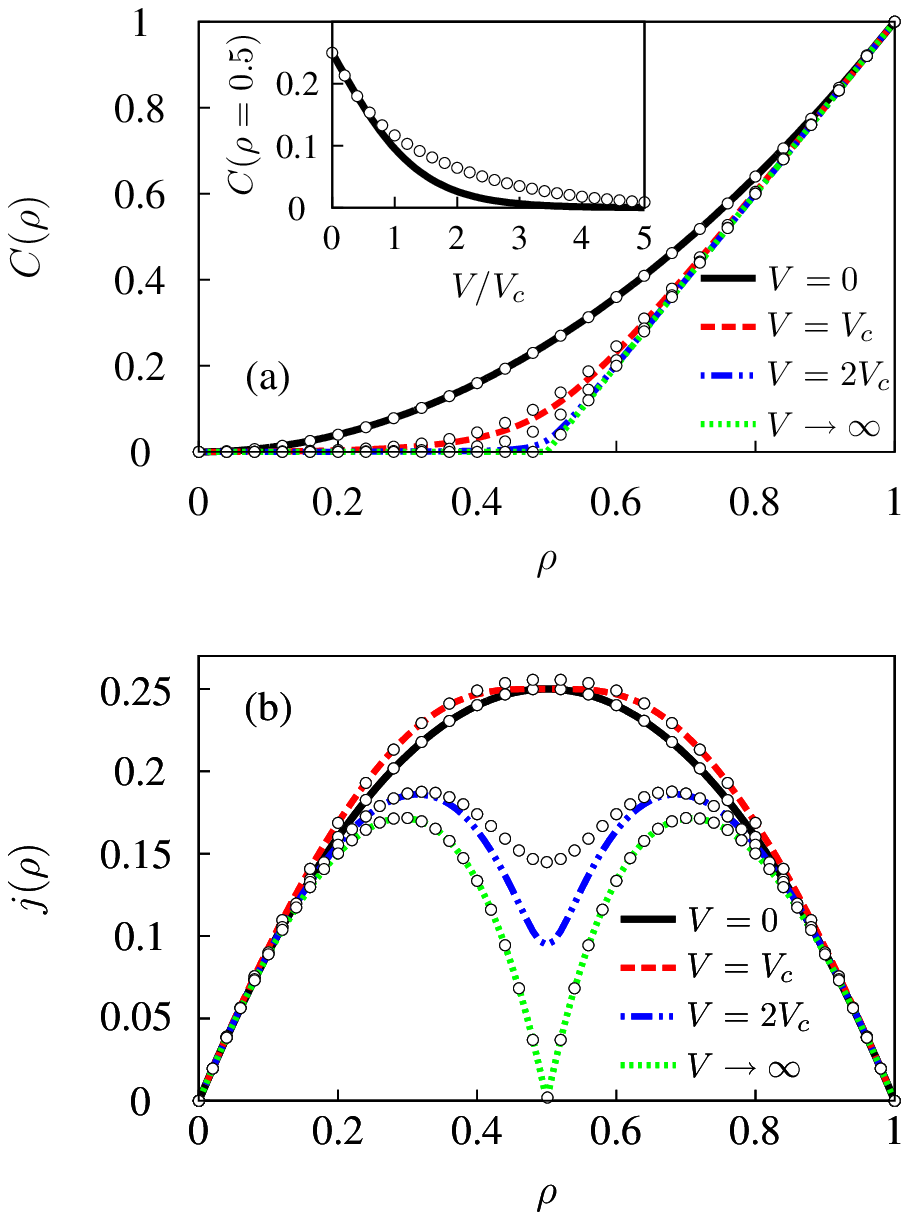}
 \caption{(Color online) (a) Two-point correlator $C(\rho)$ and (b)
   current-density relation $j(\rho)$ for various $V$ calculated by
   Eqs.~\eqref{eq:current-tdft} and \eqref{eq:correlator} (lines). The
   inset in Fig.~\ref{fig:fig1}(a) shows the pair correlator as a
   function of $V/V_{\rm c}$ for fixed density $\rho = 0.5$. The
   analytical results are compared with the true non-equilibrium
   quantities obtained by Monte Carlo simulations (circles) for a ring
   system with $1000$ sites. Each data point results from an average
   over a time corresponding to $10^9$ particle jumps in the steady
   state.}
 \label{fig:fig1}
\end{figure}
%-----------------------------------------------------------------

In order to express the correlators in Eq.~(\ref{eq:current2}) in
terms of the densities, we now apply the TDFT, which is based on the
(time-)local equilibrium approximation. According to this,
$P(\boldsymbol{n},t)$ is expressed by the Boltzmann probability
associated with $\mathcal{H}(\boldsymbol{n})$ plus a time-dependent
external potential $U[\boldsymbol{n},\boldsymbol{\rho}(t)]=
\sum_iv_i[\boldsymbol{\rho}(t)]n_i$ that equals the potential which
would generate $\boldsymbol{\rho}(t)=\lbrace\rho_i(t)\rbrace$ as
equilibrium density profile (according to the classical version of the
Mermin theorem this potential is unique for given interaction). In
effect this implies that the correlators at any time $t$ are supposed
to be related to the densities as in an equilibrium system.

To find the equilibrium correlator-density relations, we
apply the methods developed in the Markovian approach to derive exact
density functionals \cite{Buschle:2000}. Accordingly we express the
joint probabilities $p^{(j+1)}_\text{eq}(n_i,\ldots,n_{i+j})$ for the
occupation numbers $n_i,\ldots,n_{i+j}$ in equilibrium by the Markov
chain $p^{(j+1)}_\text{eq}(n_i,\ldots,n_{i+j})=p^{(1)}_\text{eq}(n_i)
\prod_{s=1}^{j}w(n_{i+s}|n_{i+s-1})$, where $p^{(1)}_\text{eq}(n_i)$
is the equilibrium probability for $n_i$, and $w(n_{i+1}|n_i)=
p^{(2)}_\text{eq}(n_i,n_{i+1})/p^{(1)}_\text{eq}(n_i)$ the conditional
probability for $n_{i+1}$ if $n_i$ is given. Since the joint
probabilities are directly connected to the correlators, e.~g.,
$p^{(4)}_\text{eq}(n_{i-1}\!=\!0,n_i\!=\!1,n_{i+1}\!=\!0,n_{i+2}\!=\!1)
=\langle\tilde n_{i-1}n_i\tilde n_{i+1}n_{i+2}\rangle_\text{eq}$, all
correlators involving more than two occupation numbers in
Eq.~(\ref{eq:current2}) can thus be reduced to two-point correlators.
The TDFT expression for the current hence becomes
\begin{align}
\label{eq:current-tdft}
j_i^{\rm\scriptscriptstyle
  TDFT}&=\left[(\rho_{i+2} - C_{i+1})e^{-V/2} + \tilde \rho_{i+1} -
 \rho_{i+2} + C_{i+1}\right] \frac{\rho_i - C_i}{\rho_i \tilde
  \rho_{i+1}}\left[\rho_i - C_{i-1} + e^{V/2}C_{i-1}\right]\,,
\end{align}
with $\tilde \rho_i=1-\rho_i$ and
$C_i=\langle n_i n_{i+1}\rangle_t$.  The two-point correlators are
related to the densities via \cite{Buschle:2000}
\begin{equation}
  C_i = \exp\left(-V\right) \frac{\left(\rho_i- C_i\right)
  \left(\rho_{i+1}- C_i\right) } {1 - \rho_i - \rho_{i+1} + C_i}\,,
  \label{eq:correlator}
\end{equation}
which can be explicitly solved to yield functions
$C_i=C_i(\rho_i,\rho_{i+1})$. In this way, the currents
$j_i^{\rm\scriptscriptstyle TDFT}$ are given as functionals
of the density profile $\boldsymbol{\rho}(t)$.

To test the quality of the TDFT approach we start by considering a
homogeneous system with periodic boundary conditions. In this case we
suppress the site indices in Eq.~\eqref{eq:correlator} and set
$j_i=j(\rho)$ in Eq.~\eqref{eq:current-tdft}. For $V\to0$,
$j(\rho)$ approaches the parabola $j =\rho-\rho^2$ for particles
feeling only hard-core repulsion. When $V$ exceeds a critical value
$V_{\rm c}=-2\ln\left(\sqrt{5}-2\right)\simeq2.89$, $j(\rho)$ develops
a double-hump structure with two maxima at densities
$\rho_{1,2}^*=\rho_{1,2}^*(V)$. In the limit $V\to\infty$ we find
$\rho^*_{1,2}=1/2\pm\left(\sqrt{2}-1\right)/2$, in agreement with earlier findings
reported by Krug \cite{Krug:1991}, and
$j\to(x^{3/2}-2x+x^{1/2})/(1-x)$ with $x=(2\rho-1)^2$, meaning that
the particle movement is frozen for a half-filling system.

In Fig.~\ref{fig:fig1} we compare the analytical findings (a) for the
pair correlator $C(\rho)$ and (b) the current-density relation
$j(\rho)$ with results from KMC simulations for various $V$. For
$V\lesssim V_{\rm c}$ as well as for large $V\gtrsim4V_{\rm c}$, we
find an excellent agreement with the simulation data, see also the
inset in Fig.~\ref{fig:fig1}(a), which shows the interaction
dependence of $C(\rho=0.5)$. For intermediate interaction strength
$V_{\rm c}<V<4V_{\rm c}$ and near half-filling, some deviations
occur. An interesting effect is seen in Fig.~\ref{fig:fig1}(b) for
weak couplings $V\lesssim V_{\rm c}$: both the theoretical predictions
and the simulations show an increased current compared to the case
$V=0$. This phenomenon is caused by effective particle-hole attraction
for weak coupling strengths.

%------ FIGURE 2 ------------------------------------------------
\begin{figure}[H!]
\centering
 \includegraphics[width=0.6\textwidth]{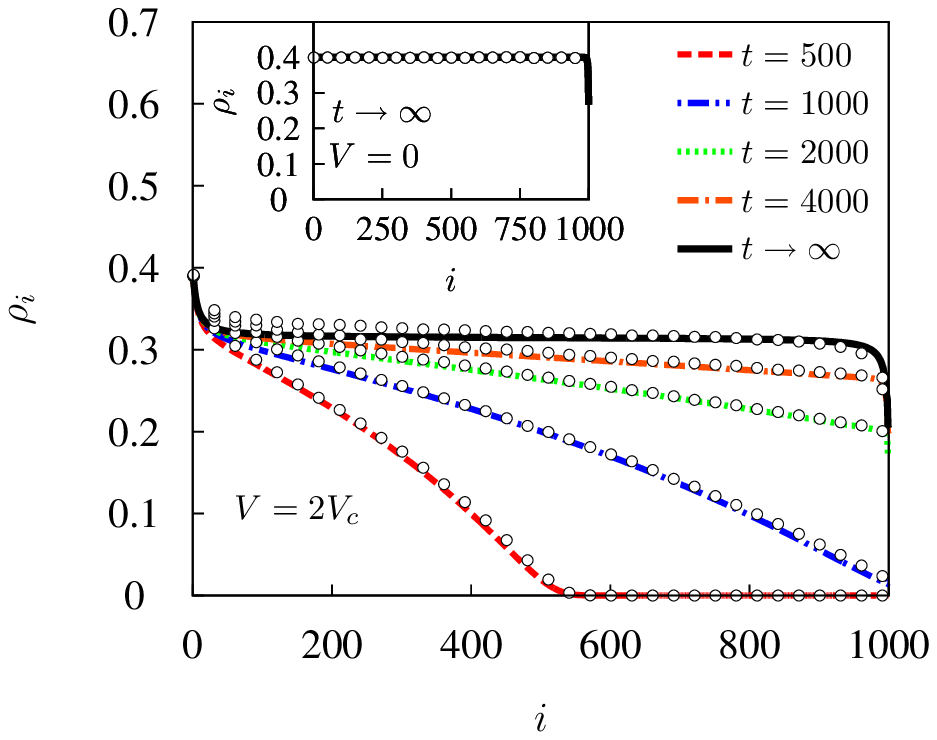}
 \caption{(Color online) Time evolution of $\rho_i$ from an initially
   empty lattice with $\rho_\text{L} = 0.4$, $\rho_\text{R} = 0.1$, $V
   = 2V_{\rm c}$ and $N = 1000$. TDFT solutions are marked by
   lines. The KMC results (circles) are averaged over $10^6$
   configurations. For $V=0$, the steady-state density profile is
   shown in the inset.}
 \label{fig:fig2}
\end{figure}
%-----------------------------------------------------------------

Next we consider an open system of $N$ sites that is coupled to two
particle reservoirs L and R at the left and right boundaries with
densities $\rho_{\rm L}$ and $\rho_{\rm R}$, respectively. The
dynamics of injection (ejection) of particles from the left (right)
reservoir is defined such that the same correlator-density relations
apply at the boundaries as in the corresponding bulk (periodic ring)
systems with densities $\rho_{\rm L}$ and $\rho_{\rm R}$. We used the
method described in \cite{Hager:2001} to implement this model in the
KMC simulations.

Let us first see how well the TDFT approach captures the time
evolution of density profiles. To this end we consider reservoir
densities $\rho_\text{L}=0.4$ and $\rho_\text{R}=0.1$ and an initially
empty lattice system at time $t=0$. For the interaction we choose
deliberately $V=2V_{\rm c}$, where comparatively large deviations were
seen in Fig~\ref{fig:fig1}. Such a ``worst-case'' choice permits the
best evaluation of approximation limits. In Fig.~\ref{fig:fig2} we
compare the results from numerical solutions of the nonlinear rate
equations (\ref{eq:rate_eq}) [where the $j_i$ are given by
Eqs.~(\ref{eq:current-tdft}) and (\ref{eq:correlator})] with the KMC
results for four different times and in the stationary limit
$t\to\infty$. The excellent agreement between the two data sets
demonstrates the power of our TDFT approach for driven lattice gas
systems with interactions. In the inset of Fig.~\ref{fig:fig2} we show
the steady state ($t\to\infty$) profile for the corresponding
non-interacting case $V=0$. As one would expect from an application of
the maximum and minimum current principle to the bulk current-density
relation shown in Fig.~\ref{fig:fig1}(b), different bulk densities
$\rho_{\rm B}$ are found for the two different interaction strengths:
$\rho_{\rm B}=\rho_{\rm L}=0.4$ for $V=0$ and $\rho_{\rm
  B}=\rho_1^*(2V_{\rm c})\simeq0.31$ for $V=2V_{\rm c}$.

\begin{figure}[H!]
\centering
 \includegraphics[width=0.6\textwidth]{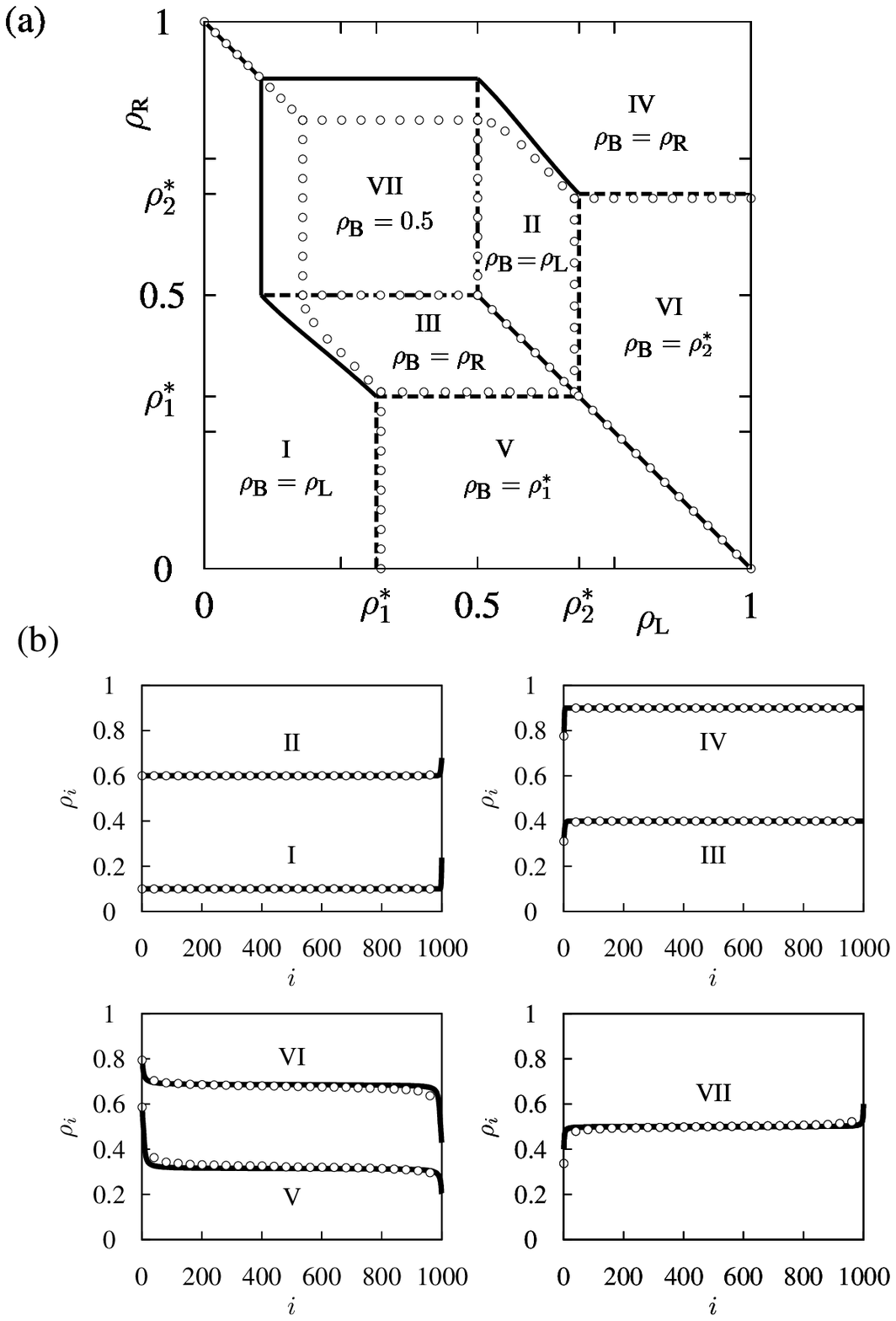}
 \caption{(a) Steady-state phase diagram and (b) density profiles of a
   system with $1000$ sites and $V = 2V_{\rm c}$. The phase diagram contains
   seven phases which are represented by density profiles
   corresponding to I ($\rho_\text{B} = \rho_\text{L}$, low-density
   phase): $\rho_\text{L} = 0.1$, $\rho_\text{R} = 0.4$; II
   ($\rho_\text{B} = \rho_\text{L}$): $\rho_\text{L} = 0.6$,
   $\rho_\text{R} = 0.7$; III ($\rho_\text{B} = \rho_\text{R}$):
   $\rho_\text{L} = 0.3$, $\rho_\text{R} = 0.4$; IV ($\rho_\text{B} =
   \rho_\text{R}$, high-density phase: $\rho_\text{L} = 0.6$,
   $\rho_\text{R} = 0.9$; V ($\rho_\text{B} = \rho_1^*$,
   maximal-current phase): $\rho_\text{L} = 0.6$, $\rho_\text{R} =
   0.1$; VI ($\rho_\text{B} = \rho_2^*$, maximal-current phase):
   $\rho_\text{L} = 0.9$, $\rho_\text{R} = 0.4$; VII ($\rho_\text{B} =
   0.5$, minimal-current phase): $\rho_\text{L} = 0.3$, $\rho_\text{R}
   = 0.7$. Solid and dashed lines in the phase diagram, which denote
   first- and second-order phase transitions, respectively, as well as
   density profiles are calculated using TDFT. Data points are
   obtained by Monte Carlo simulations with the same statistics as in
   Fig.~\ref{fig:fig1}.  }
 \label{fig:fig3}
\end{figure}

More generally, using the minimum and maximum current principle, we
can evaluate the phase diagram of the NESS for arbitrary interaction
strengths based on the bulk current-density relation derived above.
Alternatively we can use the $t\to\infty$ limit of the rate equations
to identify the bulk densities of the NESS. As expected, we find that
the results of both procedures agree \cite{comm:boundary-couplings}.
As an example we show in Fig.~\ref{fig:fig3}(a) the phase diagram of
the NESS for $V=2V_{\rm c}$, and for each of the occurring seven
phases we display the stationary density profile in
Fig.~\ref{fig:fig3}(b). In all cases the predictions are well
confirmed by the KMC results. The largest deviations occur for the
transition lines between the phases I/III, I/VII and the ``mirror
lines'' IV/II and IV/VII. Note that the phase diagram in
Fig.~\ref{fig:fig3}(a) exhibits particle-hole symmetry as required by
the Hamiltonian in Eq.~(\ref{eq:Hamiltonian}) and the dynamics
specified in Eq.~(\ref{eq:rate}). There are two maximal current phases
V and VI characterized by the bulk densities $\rho_{\rm B}=\rho_1^*$
and $\rho_{\rm B}=\rho_2^*$, respectively, and one minimal current
phase with $\rho_{\rm B}=0.5$. The remaining four phases are
determined by the reservoir densities $\rho_\text{L}$ and
$\rho_\text{R}$.

%------ FIGURE 3 ------------------------------------------------
%-----------------------------------------------------------------

In summary, we presented a new analytical approach to describe the
kinetics of driven lattice gas systems with interactions beyond
hard-core exclusions based on the TDFT. The approach was demonstrated
for a TASEP with nearest-neighbor interaction and provided very good
results for the kinetic behavior. It is clear that the TDFT is not
restricted to the TASEP situation or nearest-neighbor interactions,
but can be applied also to ASEPs (or purely boundary driven systems)
as well as other types of interactions. In the TDFT closed nonlinear
evolution equations for the densities result due to local relations
between correlators and densities, which allow one to capture the
kinetics much better than simple factorization schemes (note that
simple factorization schemes of correlators would completely fail for
$V>V_{\rm c}$ in the example studied in this work). Although one can
regard it as a weakness, it is, on the hand, advantageous that the
TDFT relies on density functionals for equilibrium systems. The
development and improvement of such equilibrium functionals in
condensed matter systems have been an intensive research area in the
past, in particular also with respect to find good approximation
schemes for the setup of functionals in dimensions larger than
one. Hence one can expect that these developments will be useful to
treat more complex non-equilibrium systems. We hope that our
findings will stimulate further research in this direction.

We thank W.~Dieterich for very valuable discussions.


\begin{thebibliography}{99}

\bibitem{Schmittmann:1995} B.\ Schmittmann and R.\ K.\ P.\ Zia, in
  {\it Phase Transitions and Critical Phenomena}, edited by
  C.\ Domb and J.\ L.\ Lebowitz (Academic Press, London, 1995), Vol.\ 17.

\bibitem{Derrida:1998} B.\ Derrida, Phys.\ Rep.\ \textbf{301}, 65
  (1998).

\bibitem{Schuetz:2001} G.\ M.\ Sch\"utz, in {\it Phase Transitions and
    Critical Phenomena}, edited by C.\ Domb and J.\ L.\
  Lebowitz (Academic Press, San Diego, 2001), Vol.\ 19.

\bibitem{Golinelli:2006} O.\ Golinelli and K.\ Mallick, J.\ Phys.\ A
  \textbf{39}, 12679 (2006).

\bibitem{Blythe:2007} R.\ A.\ Blythe and M.\ R.\ Evans, J.\ Phys.\ A
  \textbf{40}, R333 (2007).

\bibitem{MacDonald:1968} C.\ T.\ MacDonald, J.\ H.\ Gibbs, and A.\ C.\
  Pipkin, Biopolymers \textbf{6}, 1 (1968).

\bibitem{Lipowsky:2001} R.\ Lipowsky, S.\ Klumpp, and T.\ M.\
  Nieuwenhuizen, Phys.\ Rev.\ Lett.\ \textbf{87}, 108101 (2001).

\bibitem{Nishinari:2005} K.\ Nishinari, Y.\ Okada, A.\ Schadschneider,
  and D.\ Chowdhury, Phys.\ Rev.\ Lett.\ \textbf{95}, 118101 (2005).

\bibitem{Hille:2001} B.\ Hille, {\it Ionic Channels of Excitable
    Membranes}, 3rd ed.\ (Sinauer Associates, Sunderland, MA, 2001).

\bibitem{Einax:2010} M.\ Einax, M.\ K\"orner, P.\ Maass, and A.\
  Nitzan, Phys.\ Chem.\ Chem.\ Phys.\ \textbf{12}, 645 (2010).

\bibitem{Joachim:2005} C.\ Joachim and M.\ A.\ Ratner, Proc.\ Natl.\
  Acad.\ Sci.\ USA \textbf{102}, 8801 (2005).

\bibitem{Spohn:1991} J.\ Krug and H.\ Spohn, in {\it Solids far from
    Equilibrium}, edited by C.\ Godr\`eche (Cambridge University
  Press, Cambridge, 1991).

\bibitem{Helbing:2001} D.\ Helbing, Rev.\ Mod.\ Phys.\ \textbf{73},
  1067 (2001).

\bibitem{Krug:1991} J.\ Krug, Phys.\ Rev.\ Lett.\ \textbf{67}, 1882
  (1991).

\bibitem{Derrida:1992} B.\ Derrida, E.\ Domany, and D.\ Mukamel, J.\
  Stat.\ Phys.\ \textbf{69}, 667 (1992).

\bibitem{Lakatos:2003} G.\ Lakatos and T.\ Chou, J.\ Phys.\ A
  \textbf{36}, 2027 (2003).

\bibitem{Frey:2003} A.\ Parmeggiani, T.\ Franosch, and E.\ Frey,
  Phys.\ Rev.\ Lett.\ \textbf{90}, 086601 (2003).

\bibitem{Katz:1984} S.\ Katz, J.\ L.\ Lebowitz, and H.\ Spohn, J.\
  Stat.\ Phys.\ \textbf{34}, 497 (1984).

\bibitem{Hager:2001} J.\ S.\ Hager, J.\ Krug, V.\ Popkov, and G.\ M.\
  Sch\"utz, Phys.\ Rev.\ E \textbf{63}, 056110 (2001).

\bibitem{Gouyet:2003} J.-F.\ Gouyet, M.\ Plapp, W.\ Dieterich, and P.\
  Maass, Adv.\ Phys.\ \textbf{52}, 523 (2003).

\bibitem{Buschle:2000} J.\ Buschle, P.\ Maass, and W.\ Dieterich, J.\
  Phys.\ A \textbf{33}, L41 (2000).

\bibitem{comm:rates} This choice is motivated by the fact that these
  rates agree with the forward rates of an ASEP, which fulfills
  detailed balance conditions with respect to the equilibrium
  Boltzmann distribution in a non-driven system.

\bibitem{comm:boundary-couplings} We note that the TDFT allows one to
  treat general boundary couplings, where the correlator-density
  relations at the system boundaries can be different from those in
  the bulk. In this case the maximum/minimum current principle cannot
  be applied generally since the stationary density profiles become
  non-monotonous. This, however, is not the focus of the present
  work and will be discussed in detail elsewhere.

\end{thebibliography}
\end{document}